# Quantifying the amplitudes of ultrafast magnetization fluctuations in $Sm_{0.7}Er_{0.3}FeO_3$ using femtosecond noise correlation spectroscopy

M. A. Weiss[1,*], F. S. Herbst[1], G. Skobjin[1], S. Eggert[1], M. Nakajima[2], D. Reustlen[1], A. Leitenstorfer[1], S. T. B. Goennenwein[1] & T. Kurihara[1,3,†]

[1]Department of Physics, University of Konstanz, D-78457 Konstanz, Germany
[2]Institute of Laser Engineering, Osaka University, 565-0871 Osaka, Japan
[3]Department of Basic Science, The University of Tokyo, 153-8902 Tokyo, Japan

Spin fluctuations are an important issue for the design and operation of future spintronic devices. Femtosecond noise correlation spectroscopy (FemNoC) was recently applied to detect ultrafast magnetization fluctuations. FemNoC gives direct access to the spontaneous fluctuations of the magnetization in magnetically ordered materials. In FemNoC experiments, the magnetic fluctuations are imprinted on the polarization state of two independent femtosecond probe pulses upon transmission through a magnetic sample. Using a subharmonic demodulation scheme, the cross-correlation of the signals from both pulse trains is calculated. Here, we quantitatively link the FemNoC output signal to an optical polarization rotation, and then in turn to the magnitude of the inherent spin fluctuations. To this end, three different calibration protocols are presented and compared in accuracy. Ultimately, we quantitatively determine both the variance of optical polarization noise in $rad^2$, and that of the ultrafast magnetization fluctuations in $(A/m)^2$.

## I. INTRODUCTION

Macroscopic noise in condensed-matter systems often originates from microscopic fluctuations [1]. For example, the transport of charge by quantized units produces large-scale electronic shot noise [2]. Also, fluctuations of elementary spins cause macroscopic magnetization noise as seen for example in paramagnets [3–6]. Studying such phenomena provides valuable insights into the functionality of physical systems. In particular, the dynamical noise properties reveal the timescales of elementary excitations. Optical spin noise spectroscopy (SNS) [3–13] enables such investigations. In SNS, spin fluctuations are transferred to the polarization state of a laser via the Faraday effect, which may be analyzed in the megahertz to gigahertz range [6,13]. Physical information such as the paramagnetic Larmor frequency and the spin relaxation times $T_1$ and $T_2$ (Ref. [3]), homogeneous and inhomogeneous linewidths in quantum dot ensembles [12], and the spatially resolved doping concentrations in semiconductors [9] may be accessed by such measurements.

Recently, there has been growing interest in noise correlations at terahertz frequencies [14,15]. For example, the spin excitations in antiferromagnets fall in this range. Due to their high-frequency magnons, antiferromagnets are promising materials for future ultrafast spintronic devices. To study spin fluctuations at such high frequencies, femtosecond noise correlation spectroscopy (FemNoC) [16–18] based on a subharmonic detection scheme has recently been developed. FemNoC uses two ultrashort probe pulses, which are transmitted through the magnetic sample with variable time delays. Upon transmission, transient magnetization noise is translated into polarization noise by the Faraday effect. This Faraday rotation (FR) noise is then analyzed in two separate detection branches, subharmonically demodulated and finally cross-correlated in a lock-in amplifier.

However, directly linking the correlated lock-in output to quantifiable FR noise is challenging due to the stochastic nature of the measurement scheme. Thus far, FemNoC has only enabled qualitative noise evaluation. In this work, we present three calibration methods to quantify the lock-in output as effective FR noise. We compare their accuracy and extract a combined scaling factor to calibrate the lock-in output in units of $rad^2$. Ultimately, this procedure allows us to quantify the magnitude of ultrafast magnetization fluctuations in $Sm_{0.7}Er_{0.3}FeO_3$ using its magneto-optic constant.

## II. METHODS

The experimental scheme is depicted in FIG 1. Two 40 MHz probe pulse trains with variable time delay $\Delta t$ are focused onto a magnetically ordered sample using a microscope objective lens (20x magnification). They consist of femtosecond laser pulses with 300 fs duration and wavelengths centered around 767 nm and 775 nm , respectively. They are generated by spectrally and spatially separating the frequency-doubled output of a mode-locked Er.fiber laser (fundamental wavelength: 1.55 µm, pulse energy: 5nJ, repetition rate: 40 MHz), using a dichroic mirror. Furthermore, they are linearly polarized under an angle of $45°$ with respect to the optical table.

Upon transmission through the magnetic sample, the transient out-of-plane (oop) magnetization noise $\delta M_z(t)$ along the laser propagation direction induces polarization noise $\delta\alpha_{1,2}(t)$ in the probe beams 1 and 2 via the Faraday effect. This results in the time-varying polarization:

$$\alpha_{1,2}(t) = \langle\alpha_{1,2}\rangle + \delta\alpha_{1,2}(t) \qquad (1)$$

Here, $\langle\alpha_{1,2}\rangle$ is the mean polarization, periodic with the 40 MHz repetition rate $f_{rep}$ of the femtosecond pulse train. The beams are guided into two separate polarimetric detection branches using a dichroic mirror. Each branch contains a half-wave plate (HWP), a Wollaston prism (WP) and balanced photodetectors (BPD). The WP separates the probe beams into s- and p-polarized components, which are subsequently focused on the BPD. Here, a output voltage proportional to the optical power difference $P_{1,2_s} - P_{1,2_p}$ of s- and p-polarized components, the gain $G$ of the photo amplifier, and the wavelength $\lambda$ dependent responsivity $R(\lambda)$ is generated [19]. For high-frequency signals, the output voltage also needs to be scaled with the frequency $f$ and gain $G$ dependent spectral response $s(G,f)$ of the photodetector. This yields the final output:

$$V_{BPD_{1,2}} = R(\lambda)s(G,f)\,G \cdot \left(P_{1,2_s} - P_{1,2_p}\right)$$

(2)

*Contact author: marvin.weiss@uni-konstanz.de

†Contact author: takayuki.kurihara@issp.u-tokyo.ac.jp

The power difference depends on the polarization angle $\alpha_{1,2}$ and mean power $P_0$ of the probe [20]:

$$\begin{aligned}\left(P_{1,2_s} - P_{1,2_p}\right) &= P_0\big(1 - 2\cos^2(\alpha_{1,2})\big) \\ &= P_0\big(1 - 2\cos^2(\langle\alpha_{1,2}\rangle + \delta\alpha_{1,2})\big)\end{aligned}$$

(3)

Combining equations (2) and (3) gives:

$$V_{BPD_{1,2}} = R(\lambda)s(G,f)G \cdot P_0\big(1 - 2\cos^2(\langle\alpha_{1,2}\rangle + \delta\alpha_{1,2})\big)$$

(4)

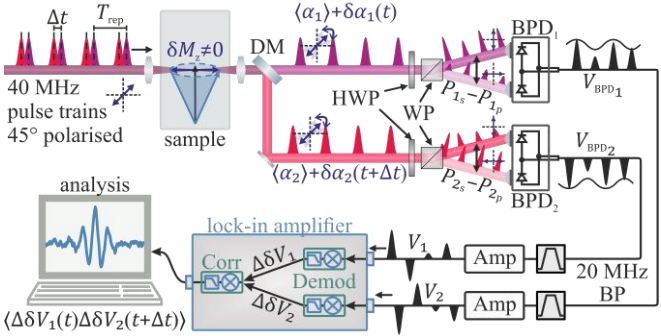

FIG 1. Overview of the experimental setup and electronic data processing. Two femtosecond laser pulse trains with marginally different wavelength (center wavelengths of 767 nm (purple) and 775 nm (red)) and variable time delay $\Delta t$ experience a polarization rotation $\delta\alpha_{1,2}$ proportional to out-of-plane spin fluctuations $\delta M_z$ in addition to their initial polarization angle of $\langle\alpha_{1,2}\rangle = 45°$ (in respect to the optical table). $\delta\alpha_{1,2}$ are measured in separate detection branches, each consisting of a half-wave plate (HWP), a Wollaston prism (WP), and a balanced photodetector (BPD). The WP splits the light into s- and p-polarized components (blue arrows) with optical powers $P_{1,2_s}$ and $P_{1,2_p}$, respectively. The BPDs output voltages $V_{BPD_{1,2}}$ proportional to $P_{1,2_s} - P_{1,2_p}$. After amplification and extraction of the pulse-to-pulse voltage fluctuations $\Delta\delta V_{1,2}$ from each branch, the cross-correlation function $\langle\Delta\delta V_1(t)\Delta\delta V_2(t+\Delta t)\rangle$ is calculated in real time. DM: dichroic mirror; BP: electronic bandpass filter; Amp: amplifier; Demod: demodulator; Corr: real-time correlation.

It shows that the BPD output reflects both the mean polarization $\langle\alpha\rangle$ and fluctuations $\delta\alpha$. Note that $V_{BPD_{1,2}}$ is dominated by $\langle\alpha_{1,2}\rangle$ periodic with the repetition rate $f_{rep} = 40$ MHz, whereas the noise is mostly contained in the first subharmonic frequency $\frac{f_{rep}}{2} = 20$ MHz (pulse-to-pulse fluctuations) [18,21]. The HWPs compensate for any additional static polarization components $\langle\alpha_{1,2}\rangle$, ensuring that $\langle\alpha_{1,2}\rangle \approx 45°$. If this condition is fulfilled and fluctuations are small ($\delta\alpha_{1,2} \approx 0$) the optical power difference simplifies to [18]:

$$\left(P_{1,2_s} - P_{1,2_p}\right) = 2P_0\delta\alpha_{1,2}\frac{1}{\text{rad}}$$

(5)

Note that the dimensionality factor $\frac{1}{\text{rad}}$ appears due to the small-angle approximation. This factor will be omitted in the following. Inserting equation (5) into equation (4) then yields the detector output:

$$V_{BPD_{1,2}} \approx 2P_0 R(\lambda)s(G,f)G \cdot \delta\alpha_{1,2}$$

(6)

Therefore, analyzing $V_{BPD_{1,2}}$ grants direct access to polarization fluctuations. In practice, a dominant 40 MHz component persists. We utilize a 20 MHz bandpass filter (BP) and amplifier (Amp) to enhance the subharmonic pulse-to-pulse fluctuation, and at the same time suppress the fundamental frequency. This enables full use of the lock-in amplifier's dynamic range defined by its digitization depth.

Next, the filtered voltages $V_1$ and $V_2$ are fed into demodulation channels of the lock-in amplifier (UHFLI, Zurich Instruments [22]) for subharmonic demodulation [17,18]. Here, the 20 MHz pulse-to-pulse fluctuations $\Delta\delta V_{1,2}$ are extracted from the low-frequency and fundamental laser components. Subsequently, $\Delta\delta V_1$ and $\Delta\delta V_2$ are multiplied and averaged in real-time. This yields the correlated noise $\langle\Delta\delta V_1(t) \cdot \Delta\delta V_2(t+\Delta t)\rangle$ as the final output of the lock-in amplifier (see inset in FIG 1). $\langle\Delta\delta V_1(t) \cdot \Delta\delta V_2(t+\Delta t)\rangle$ is proportional to the correlated FR noise $\langle\Delta\delta\alpha_1(t) \cdot \Delta\delta\alpha_2(t+\Delta t)\rangle$ and the ultrafast magnetization fluctuation autocorrelation $\langle\delta M_z(t) \cdot \delta M_z(t+\Delta t)\rangle$ along the laser propagation:

$$\begin{aligned}\langle\Delta\delta V_1(t) \cdot \Delta\delta V_2(t+\Delta t)\rangle &= C_\alpha \cdot \langle\Delta\delta\alpha_1(t) \cdot \Delta\delta\alpha_2(t+\Delta t)\rangle \\ &= C_\alpha \cdot C_M \cdot \langle\delta M_z(t) \cdot \delta M_z(t+\Delta t)\rangle\end{aligned}$$

(7)

Here, $C_\alpha$ and $C_M$ are proportionality constants. Note that throughout this article, $\langle\Delta\delta V_1(t) \cdot \Delta\delta V_2(t+\Delta t)\rangle$ is given in the unit of [V] instead of [V$^2$] due to an internal scaling factor of $\frac{1}{V}$ applied in the lock-in amplifier [22].

The mean value of the polarization rotation $\langle\alpha\rangle$, and with that the mean oop magnetization $M_z$, can easily be determined with simple polarimetric measurement schemes [4]. In contrast, the fluctuations $\delta\alpha$ are orders of magnitude smaller and governed by different statistics. This fact significantly complicates their quantitative analysis. Nevertheless, we demonstrate three different protocols to calibrate the sample induced FR noise by calculation of the proportionality factor $C_\alpha$. Subsequently, the magneto-optic constant of $Sm_{0.7}Er_{0.3}FeO_3$ at the probing wavelength is determined to obtain $C_M$. With this information, we determine the variance of the spontaneous spin fluctuations in absolute units (A/m)$^2$.

*Contact author: marvin.weiss@uni-konstanz.de
†Contact author: takayuki.kurihara@issp.u-tokyo.ac.jp

## III. RESULTS AND DISCUSSION

### A. Faraday rotation noise calibration schemes

The first calibration method involves analyzing each signal processing step in the detection chain individually. The second method compares the correlated noise to the optical shot noise. In the third approach, an acousto-optic modulator (AOM) is used to introduce amplitude modulation of the pulses at the subharmonic frequency, which is compared to the FR noise.

#### 1. Calibration via transfer functions

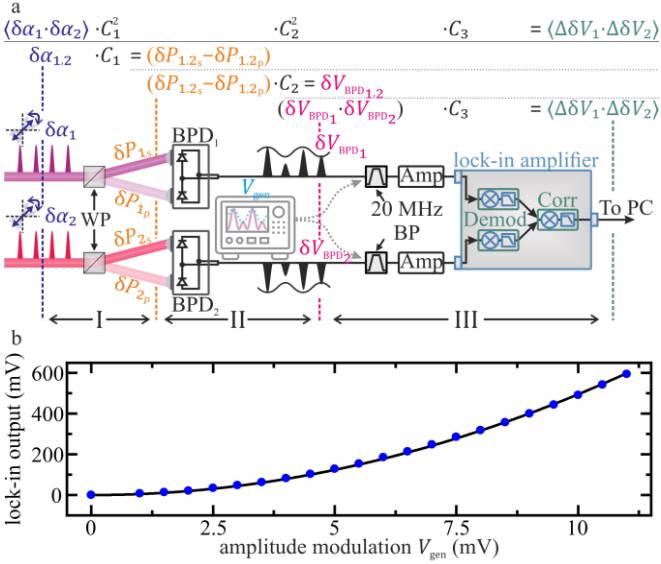

FIG 2. Calibration of the detection chain using transfer functions. a A schematic of the calibration process. Area I: After transmission through the magnetic sample, the femtosecond probe pulses exhibit polarization fluctuations $\delta\alpha_{1,2}$ relative to their mean polarization angle $\langle\alpha_{1,2}\rangle = 45°$ in respect to the optical table. A Wollaston prism (WP) splits the beams into their s- and p-polarized components. Area II: The fluctuations in the optical power are analyzed in balanced photodetectors (BPD), which output the corresponding voltage fluctuations $\delta V_{\mathrm{BPD}_{1,2}}$. Area III: $\delta V_{\mathrm{BPD}_{1,2}}$ are first bandpass filtered and amplified before subharmonic demodulation (Demod) in a lock-in amplifier. Subsequent correlation analysis (Corr) yields the correlated voltage fluctuations $\langle\Delta\delta V_1 \cdot \Delta\delta V_2\rangle$. The proportionality factor $C_3$ arising in this signal processing step is determined using a function generator (grey box), which produces a 40 MHz voltage $V_{\mathrm{gen}}$ with a 20 MHz modulation. BP: electronic bandpass filter; Amp: amplifier. b The measured correlated lock-in output $\langle\Delta\delta V_1 \cdot \Delta\delta V_2\rangle$ as a function of the 20 MHz modulation amplitude $V_{\mathrm{gen}}$. A quadratic fit ($\langle\Delta\delta V_1 \cdot \Delta\delta V_2\rangle = C_3 \cdot V_{\mathrm{gen}}^2$) reveals $C_3 = 4.95 \cdot 10^3 \frac{1}{\mathrm{V}}$.

In this method, we analyze each step of the signal processing chain to determine the linear proportionality factors $C_i$ that relate the input and output for each detection stage $i$. Using the transfer functions of each device, the total calibration factor $C_{\alpha,\mathrm{TF}}$ is determined as:

$$\langle\Delta\delta V_1 \cdot \Delta\delta V_2\rangle = \underbrace{C_1^2 C_2^2 C_3}_{C_{\alpha,\mathrm{TF}}} \cdot \langle\Delta\delta\alpha_1 \cdot \Delta\delta\alpha_2\rangle$$

(8)

*Contact author: marvin.weiss@uni-konstanz.de
†Contact author: takayuki.kurihara@issp.u-tokyo.ac.jp

We divide the setup into three key signal processing steps (see FIG 2a):

1. Step 1 (area I): Conversion of pulse-to-pulse polarization noise $\delta\alpha_{1,2}$ into the fluctuation of power difference between s- and p- polarized components $\left(\delta P_{1,2_\mathrm{s}} - \delta P_{1,2_\mathrm{p}}\right)$ using a Wollaston prism (WP). The first transfer function $C_1$ relating these two quantities is directly obtained from equation (5) and yields:

$$C_1 = 2P_0$$

(9)

2. Step 2 (area II): Optical detection, where the power fluctuations $\left(\delta P_{1,2_\mathrm{s}} - \delta P_{1,2_\mathrm{p}}\right)$ are converted into voltage fluctuations $\delta V_{\mathrm{BPD}_{1,2}}$ by balanced photodetectors (BPDs). This step is summarized in the second transfer function (see equation (4)):

$$C_2 = R(\lambda)s(G,f)G$$

(10)

where, $R(\lambda)$, $s(G,f)$ and $G$ are obtained from the BPD specifications [19], or alternatively, $R(\lambda) \cdot s(G,f)$ can be measured experimentally, e.g., by measuring the shot noise from a continuous-wave laser shone onto the detectors.

3. Step 3 (area III): Filtering and amplifying the detector outputs $\delta V_{\mathrm{BPD}_{1,2}}$, followed by subharmonic demodulation and correlation in a lock-in amplifier, yield the correlated pulse-to-pulse voltage fluctuations $\langle\Delta\delta V_1 \cdot \Delta\delta V_2\rangle$. The final calibration function $C_3$ describing this detection step is determined using a signal generator. This mimics the BPD outputs by producing a 40 MHz voltage with a 20 MHz modulation as an input for both lock-in channels. The correlated lock-in output is measured as a function of the modulation amplitude. Afterwards, $C_3$ is determined by applying the following fitting function to the data (Figure 2b):

$$\langle\Delta\delta V_1 \cdot \Delta\delta V_2\rangle = C_3 \cdot V_{\mathrm{gen}}^2$$

(11)

Inserting experimental values (see Appendix E, Table 1) and assuming both detection arms to have identical characteristics, we calculate the transfer function calibration factor:

$$C_{\alpha,\mathrm{TF}} = (4.8 \pm 1.6)\frac{\mu\mathrm{V}}{(\mu\mathrm{rad})^2}$$

(12)

Details on the uncertainty analysis are provided in Appendix E.

## 2. Calibration via comparison to the optical shot noise

In this method, we calibrate the spin noise by comparing it to the level of optical shot noise. We determine the optical shot noise contribution $\langle(\Delta\delta V)^2\rangle_{\text{SN}}$ to the total voltage noise in each detection branch. Afterwards, we relate it to the shot noise-induced fluctuations in the optical power difference of s- and p-polarized components $\langle(\delta P_{1,2_s} - \delta P_{1,2_p})^2\rangle$. From Appendix A, equation (A9), this is found to be identical to the shot noise level $\langle(\delta P)^2\rangle_{\text{SN}}$ of a single beam with optical power $P_0$:

$$\langle(\delta P_{1,2_s} - \delta P_{1,2_p})^2\rangle = \langle(\delta P)^2\rangle_{\text{SN}} = 2h\nu P_0 \Delta f$$

(13)

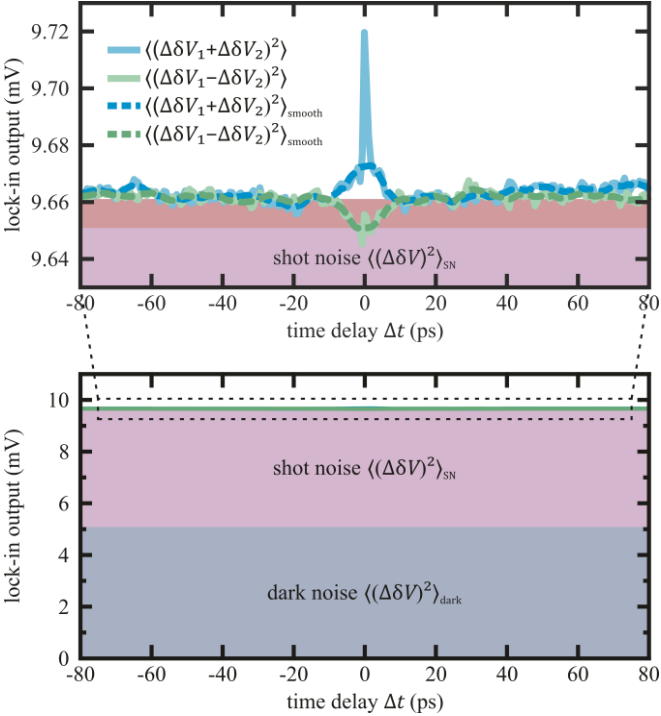

FIG 3. Calibration based on optical shot noise evaluation. Slope-corrected (solid lines) and smoothed (dotted lines) averages of the squared sum (blue) and difference (green) of pulse-to-pulse voltage fluctuations are plotted as a function of time delay $\Delta t$. A sharp peak at $\Delta t = 0$ appears in the non-smoothed curves – likely due to spectral leakage of the probes. To estimate the Faraday noise amplitude, the $\Delta t = 0$ values from the smoothed curves are used. The lock-in output is primarily dominated by shot noise $\langle(\Delta\delta V)^2\rangle_{\text{SN}}$ (purple area) and dark noise $\langle(\Delta\delta V)^2\rangle_{\text{dark}}$ (dark blue area), while the much smaller Faraday noise contributions $\langle(\Delta\delta V_{1,2})^2\rangle_{\text{FR}}$ (red area) and $\langle\Delta\delta V_1 \cdot \Delta\delta V_2\rangle_{\text{FR}}$ appear on top (see top panel). Further details on data processing are provided in Appendix C.

Here, $h$ is Planck's constant, $\nu$ is the photon frequency and $\Delta f$ is the noise bandwidth.

Now, we calculate $\langle(\delta\alpha)^2\rangle_{\text{SN}}$. We define this as the effective polarization noise which would produce equivalent power difference fluctuations as the optical shot noise. The calculation follows Malus' law [20] for the case of perfectly balanced photo detectors and small fluctuations (see equation (5)):

*Contact author: marvin.weiss@uni-konstanz.de
†Contact author: takayuki.kurihara@issp.u-tokyo.ac.jp

$$\langle(\delta P_{1,2_s} - \delta P_{1,2_p})^2\rangle = 4P_0^2 \langle(\delta\alpha)^2\rangle_{\text{SN}}$$

(14)

The voltage noise and polarization noise are linearly related:

$$\langle(\Delta\delta V)^2\rangle_{\text{SN}} = C_{\alpha,\text{SN}} \cdot 2\langle(\delta\alpha)^2\rangle_{\text{SN}}$$

(15)

with the proportionality factor $C_{\alpha,\text{SN}}$. Here, the factor of 2 arises because the pulse-to-pulse polarization fluctuation is twice the variance of $\langle(\delta\alpha)^2\rangle$ (see equation (A11)):

$$\langle(\Delta\delta\alpha)^2\rangle = 2\langle(\delta\alpha)^2\rangle$$

(16)

Inserting equation (13) and (14) into equation (15) reveals the shot noise proportionality constant:

$$C_{\alpha,\text{SN}} = \frac{\langle(\Delta\delta V)^2\rangle_{\text{SN}}}{2\langle(\delta\alpha)^2\rangle_{\text{SN}}} = \frac{\langle(\Delta\delta V)^2\rangle_{\text{SN}}}{h\nu\Delta f} \cdot P_0$$

(17)

To determine $\langle(\Delta\delta V)^2\rangle_{\text{SN}}$, we measure $\langle(\Delta\delta V_1 \pm \Delta\delta V_2)^2\rangle$ as a function of the time delay $\Delta t$, where:

$$\langle(\Delta\delta V_1 \pm \Delta\delta V_2)^2\rangle \\ = \langle(\Delta\delta V_1)^2\rangle + \langle(\Delta\delta V_2)^2\rangle \\ \pm 2\langle\Delta\delta V_1 \cdot \Delta\delta V_2\rangle$$

(18)

This expression includes both the correlated noise $\langle\Delta\delta V_1 \cdot \Delta\delta V_2\rangle$ between the detection branches and uncorrelated noise in each branch $\langle(\Delta\delta V_1)^2\rangle$ and $\langle(\Delta\delta V_2)^2\rangle$. We assume the noise in each branch $\langle(\Delta\delta V_{1,2})^2\rangle$ is dominated by three contributions: Optical shot noise $\langle(\Delta\delta V_{1,2})^2\rangle_{\text{SN}} = 2C_{\alpha,\text{SN}}\langle(\Delta\delta\alpha_{1,2})^2\rangle_{\text{SN}}$, dark noise of the electrical detection chain $\langle(\Delta\delta V_{1,2})^2\rangle_{\text{dark}}$, and FR-induced voltage noise $\langle(\Delta\delta V_{1,2})^2\rangle_{\text{FR}} \propto 2C_{\alpha,\text{SN}}\langle(\Delta\delta\alpha_{1,2})^2\rangle_{\text{FR}}$. Substituting these into equation (18), we get:

$$\langle(\Delta\delta V_1 \pm \Delta\delta V_2)^2\rangle \\ = \langle(\Delta\delta V)^2\rangle_{\text{SN}} + \langle(\Delta\delta V)^2\rangle_{\text{dark}} \\ + \langle(\Delta\delta V)^2\rangle_{\text{FR}} \pm 2\langle\Delta\delta V_1 \cdot \Delta\delta V_2\rangle_{\text{FR}}$$

(19)

Here, we defined $\langle(\Delta\delta V)^2\rangle_{\text{SN}} \stackrel{\text{def}}{=} \langle(\Delta\delta V_1)^2\rangle_{\text{SN}} + \langle(\Delta\delta V_2)^2\rangle_{\text{SN}}$, $\langle(\Delta\delta V)^2\rangle_{\text{dark}} \stackrel{\text{def}}{=} \langle(\Delta\delta V_1)^2\rangle_{\text{dark}} + \langle(\Delta\delta V_2)^2\rangle_{\text{dark}}$ and $\langle(\Delta\delta V)^2\rangle_{\text{FR}} \stackrel{\text{def}}{=} \langle(\Delta\delta V_1)^2\rangle_{\text{FR}} + \langle(\Delta\delta V_2)^2\rangle_{\text{FR}}$. Furthermore, we assume that both the dark noise and shot noise as well as their cross products are uncorrelated. Note that the voltage noise correlation $\langle\Delta\delta V_1 \cdot \Delta\delta V_2\rangle_{\text{FR}} \propto \langle\Delta\delta\alpha_1 \cdot \Delta\delta\alpha_2\rangle_{\text{FR}}$ is the relevant signal measured in our FemNoC experiment.

To quantify the components given in equation ( 19 ), we first measure $\langle(\Delta\delta V_1 \pm \Delta\delta V_2)^2\rangle$ with the laser beams blocked and obtain the dark noise level $\langle(\Delta\delta V)^2\rangle_{\text{dark}} = 5.1$ mV (FIG 3). In a next step, we measure the same quantity without blocking the lasers as a function of $\Delta t$. Now, an extremum around $\Delta t = 0$ with an amplitude of approximately 11 µV appears (top panel, FIG 3). We attribute this to the correlated FR noise due to its characteristic time dependence [17]. Employing the approximation $\langle(\Delta\delta V)^2\rangle_{\text{FR}} = 2\left\langle\left(\Delta\delta V_{1,2}\right)^2\right\rangle_{\text{FR}} = 2\langle\Delta\delta V_1 \cdot \Delta\delta V_2(\Delta t = 0)\rangle_{\text{FR}}$, we now calculate the shot noise level $\langle(\Delta\delta V)^2\rangle_{\text{SN}} = 4.57$ mV with the results obtained above.

Finally, inserting the experimental values (Appendix E, Table 2), we find the shot noise calibration factor:

$$C_{\alpha,\text{SN}} = (4.8 \pm 1.2)\frac{\mu V}{(\mu\text{rad})^2}$$

( 20 )

### 3. Calibration via quantifiable amplitude noise introduced using an acousto-optic modulator

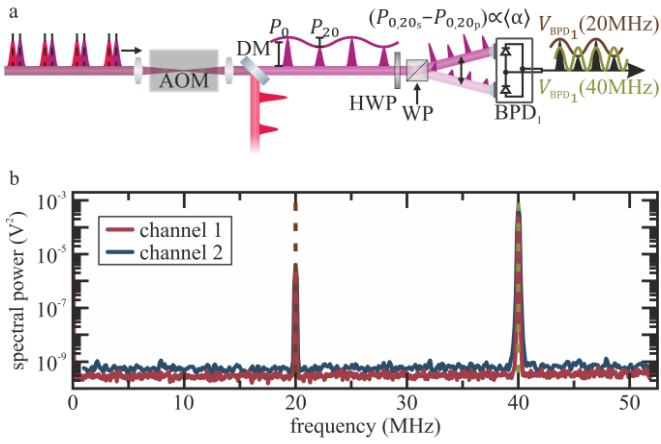

FIG 4. Calibration using an acousto-optic modulator (AOM). a Sketch of the calibration setup. In this method, the magnetic sample is replaced by an AOM, which applies a subharmonic (20 MHz) amplitude modulation with optical power $P_{20}$ on the probe beams (red and purple) with mean optical power $P_0$. The pulse trains are directed into the two detection branches via a dichroic mirror (DM). A Wollaston prism (WP) separates them into their s- and p-polarized components. Their mean polarization angle $\langle\alpha\rangle$ is adjusted using a half-wave plate (HWP). b Power spectrum of detection channel 1 (red) and 2 (blue) measured using a spectrum analyzer. Details are discussed in Appendix D.

In this third calibration method, we replace the magnetic sample with a free-space acousto-optic modulator (AOM) to artificially introduce a subharmonic ($\frac{f_{\text{rep}}}{2} = 20$ MHz) amplitude modulation on the probe beams (FIG 4a). This modulation is linked to an equivalent FR noise that would induce equivalent optical power differences after the Wollaston prism. The probe beam consists of a fundamental frequency component $f_{\text{rep}} = 40$ MHz with mean optical power $P_0$ and the 20 MHz amplitude modulation with optical power $P_{20}$. Both components induce voltages in the BPDs during polarimetric detection. According to equation ( 6 ), the voltage noise of the 20 MHz component relates to the optical power and voltage fluctuations at the same frequency. In this case, the voltage noise now depends on $\langle\alpha\rangle$ instead of the polarization fluctuations $\delta\alpha$ due to the amplitude modulation which equally affects s- and p-polarized components:

$$\left\langle V_{\text{BPD}}^2\left(\frac{f_{\text{rep}}}{2}\right)\right\rangle = s^2(G,f)G^2 \cdot \left(P_{20_s} - P_{20_p}\right)^2$$
$$\approx 4s^2(G,f)G^2 \cdot P_{20}^2\langle\alpha\rangle^2$$

( 21 )

We now aim to link this to an effective FR noise that produces equivalent subharmonic voltage noise as the amplitude modulation. Note that in this calibration method, the shot noise is much smaller than the AOM induced amplitude modulation and is thus neglected. Again, using equation ( 6 ), the voltage noise from the effective FR noise is:

$$\left\langle V_{\text{BPD}}^2\left(\frac{f_{\text{rep}}}{2}\right)\right\rangle = 4s^2(G,f)G^2 \cdot P_0^2\langle(\delta\alpha)^2\rangle$$

( 22 )

Inserting equation ( 21 ) into equation ( 22 ) gives the relationship:

$$\langle(\delta\alpha)^2\rangle = \frac{P_{20}^2}{P_0^2}\langle\alpha\rangle^2$$

( 23 )

Additionally, in this configuration, the correlation function is proportional to $\langle\alpha\rangle^2$ with proportionality factor $D$:

$$\langle\Delta\delta V_1 \Delta\delta V_2\rangle = D\langle\alpha\rangle^2 = \frac{D}{2}\frac{P_0^2}{P_{20}^2}\langle\Delta\delta\alpha_1 \Delta\delta\alpha_2\rangle$$

( 24 )

Assuming the effective FR noise in each detection branch is approximately equal, and using equation ( 16 ), the calibration factor $C_{\alpha,\text{AOM}}$ becomes:

$$C_{\alpha,\text{AOM}} = \frac{D}{2}\frac{P_0^2}{P_{20}^2}$$

( 25 )

To determine the power ratio $\frac{P_0}{P_{20}}$, we measure the BPD output at both the fundamental and subharmonic frequencies using a spectrum analyzer (FIG 4b):

$$\frac{\langle V_{\text{BPD}}^2(f_{\text{rep}})\rangle}{\left\langle V_{\text{BPD}}^2\left(\frac{f_{\text{rep}}}{2}\right)\right\rangle} = \frac{P_0^2}{P_{20}^2}$$

( 26 )

and find $\frac{\langle V_{\text{BPD}}^2(40\text{ MHz})\rangle}{\langle V_{\text{BPD}}^2(20\text{ MHz})\rangle} = 0.012$ (Appendix D). Inserting equation ( 26 ) into ( 24 ) gives the final calibration relation:

*Contact author: marvin.weiss@uni-konstanz.de
†Contact author: takayuki.kurihara@issp.u-tokyo.ac.jp

$$\langle\Delta\delta V_1 \Delta\delta V_2\rangle = \frac{D}{2}\underbrace{\frac{\langle V_{\text{BPD}}^2(f_{\text{rep}})\rangle}{\langle V_{\text{BPD}}^2\left(\frac{f_{\text{rep}}}{2}\right)\rangle}}_{C_{\alpha,\text{AOM}}}\langle\Delta\delta\alpha_1 \Delta\delta\alpha_2\rangle$$

(27)

Next, we measure $\langle\Delta\delta V_1 \Delta\delta V_2\rangle$ as a function of $\langle\alpha_1\rangle\langle\alpha_2\rangle$. Fitting the results with a polynomial function of the form $f(x) = cx + Dx^2$ reveals $D = 0.1024 \pm 0.003 \frac{\mu V}{(\mu rad)^2}$ (Appendix D, Suppl. Fig. 4).

Finally, inserting the experimental values (Appendix E, Table 3), we calculate:

$$C_{\alpha,\text{AOM}} = (4.2 \pm 1.3)\frac{\mu V}{(\mu rad)^2}$$

(28)

### B. Comparison of calibration methods and Faraday rotation sensitivity

#### 1. Comparison of calibration methods

The calibration factors determined for the three methods – transfer function (TF), shot noise (SN), and acousto-optic modulator (AOM) – are specific to the optical probing power $P_{0,\text{ref}} = 1.09$ mW at which the present calibration was performed. However, these calibration factors can be generalized for any optical power $P_0$ using the normalization:

$$C_{\alpha,\text{TF,SN,AOM}}(P_0) = \frac{P_0^2}{P_{0,\text{ref}}^2}C_{\alpha,\text{TF,SN,AOM}}(P_{0,\text{ref}})$$

(29)

To account for uncertainties in the generalization, the error in the calibration factor for any power $P_0$ is extrapolated by incorporating the uncertainties of $C_{\alpha,\text{TF,SN,AOM}}(P_{0,\text{ref}})$ into equation (29):

$$\begin{aligned}C_{\alpha,\text{TF,SN,AOM}}(P_0) &\pm u\left(C_{\alpha,\text{TF,SN,AOM}}(P_0)\right)\\ &= \frac{P_0^2}{P_{0,\text{ref}}^2}(C_{\alpha,\text{TF,SN,AOM}}(P_{0,\text{ref}}))\\ &\pm u\left(C_{\alpha,\text{TF,SN,AOM}}(P_{0,\text{ref}})\right)\end{aligned}$$

(30)

A comparison between equations (29) and (30) shows that although the absolute uncertainty of the calibration factor increases with higher probe powers, the relative uncertainty remains constant. Despite the increasing uncertainty with higher $P_0$, the three methods show comparable calibration factors at $P_0 \approx P_{0,\text{ref}}$, with all falling within their respective error margins.

To improve the accuracy of the final calibration, we utilize the inverse-variance weighted mean [23], which gives more weight to calibration values with smaller uncertainties. This approach incorporates the experimental errors of the individual methods (see Appendix E) and yields a final calibration factor:

$$C_{\alpha,\text{AVE}}(P_{0,\text{ref}}) = (4.6 \pm 0.8)\frac{\mu V}{(\mu rad)^2}$$

(31)

By using equations (29) and (30), the calibration factor and corresponding uncertainty can be extrapolated for any power $P_0$. It is important to note that the individual calibration factors $C_{\alpha,\text{TF}}$, $C_{\alpha,\text{SN}}$, and $C_{\alpha,\text{AOM}}$ carry uncertainties of approximately 25–33%. This reflects the inherent challenge of calibrating stochastic fluctuations, where standard calibration procedures do not apply. Within this context, we consider the reported accuracy to be a realistic and robust result.

#### 2. Faraday rotation sensitivity

We apply the combined calibration factor $C_{\alpha,\text{AVE}}$ to the measured correlated noise time traces. The data consists of the average of $n = 7$ correlated noise time traces recorded on a 10 μm thick sample of the antiferromagnet $Sm_{0.7}Er_{0.3}FeO_3$ kept at room temperature ($T \approx 297$ K). An external magnetic field $B = 63 \pm 5$ mT is applied in the oop direction to suppress stochastic picosecond random telegraph noise near the spin reorientation transition (Ref. [17]). Each time trace spans $\Delta t = -80$ ps to $\Delta t = +80$ ps in 1 ps steps with probing power $P_{0,\text{ref}} = 1.09$ mW and averages taken from approximately $10^7$ correlation samples per data point.

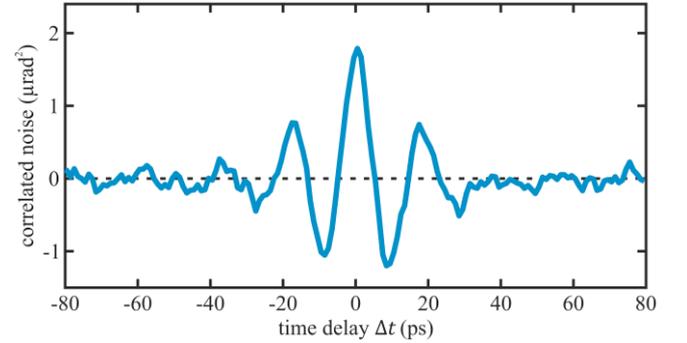

FIG 5: Calibrated correlated Faraday noise in $Sm_{0.7}Er_{0.3}FeO_3$ at $T = 297$ K and external out-of-plane field of $B = 63 \pm 5$ mT as a function of time delay $\Delta t$. The data represents the baseline-subtracted average of seven correlated noise time traces (see Appendix F for details), calibrated by dividing the raw data by the calibration factor $C_{\alpha,\text{AVE}}$.

FIG 5 shows the calibrated time-trace average, which is obtained by dividing the raw data by the calibration factor $C_{\alpha,\text{AVE}}$. The correlated FR noise exhibits a peak at $\Delta t = 0$ with an amplitude of approximately 2 $(\mu rad)^2$, followed by a symmetric damped oscillation with a characteristic frequency in the picosecond range. As discussed in Ref. [17], this oscillation directly reflects the temporal autocorrelation of thermal fluctuations in the quasiferromagnetic (qF) magnon mode of $Sm_{0.7}Er_{0.3}FeO_3$. The observed picosecond oscillation corresponds to the eigenfrequency of the qF mode, while the envelope decay arises from the finite correlation time of these fluctuations. Beyond $\pm 50$ ps, the correlated FR noise amplitude drops below the background noise level, which

*Contact author: marvin.weiss@uni-konstanz.de

†Contact author: takayuki.kurihara@issp.u-tokyo.ac.jp

limits the sensitivity of our experimental setup (see Appendix F):

$$S_{\text{rad}} = 0.52 \frac{(\mu\text{rad})^2}{\text{mW}\sqrt{n}}$$

(32)

This sensitivity means that with a probe power of 1 mW, we can resolve FR noise as small as $0.52\ (\mu\text{rad})^2$ per number of averaged time traces $n$, where each data point is the average of approximately $10^7$ correlation samples. This resolution can be improved by increasing the probe power or averaging more samples. Experimentally, however, care must be taken with large powers because optical excitation effects might eventually take place [24,25].

### C. Magnetization calibration

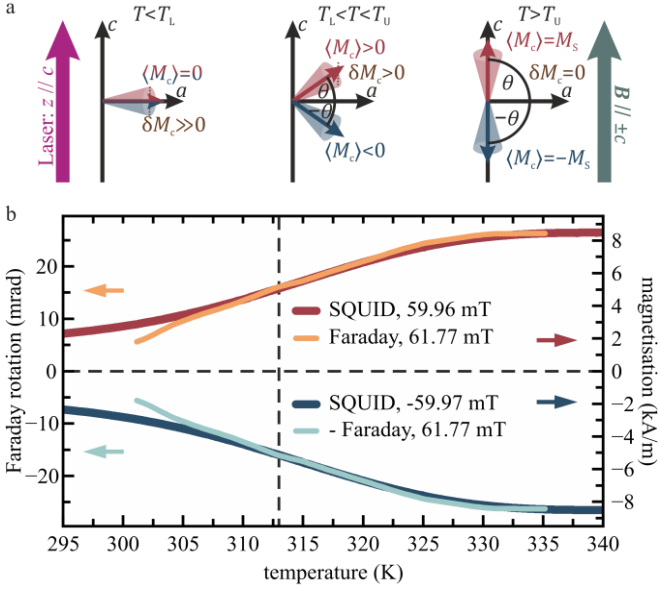

FIG 6: Faraday rotation and magnetization in $Sm_{0.7}Er_{0.3}FeO_3$. a Schematic of the spin reorientation transition (SRT). The laser and external magnetic field are aligned with the $c$-axis of the orthoferrite sample. Below the lower threshold temperature $T_L$, the net magnetization (blue and red arrows) aligns with the $a$-axis. As the SRT progresses ($T_L < T < T_U$), the magnetization rotates by an angle $\pm\theta$, which changes the c-axis projection of the mean magnetization $\langle M_c \rangle$ and fluctuation component $\delta M_c$. b Temperature-dependent Faraday rotation (FR) in comparison with magnetization data from SQUID magnetometry. To obtain the magnetization values (dark red and dark blue curves), the SQUID raw data is divided by the sample volume. The FR data is background-subtracted (see Appendix G) and shifted by +7 K to account for laser-induced heating. The light blue curve corresponds to the orange curve mirrored around the zero line.

The calibration methods presented in chapter IIIA allow quantitative identification of Faraday rotation values. It is further possible to connect this information to the absolute value of magnetization fluctuation in units of $\frac{A^2}{m^2}$. To do this, we determine the proportionality factor $C_M$, which relates correlated FR noise to inherent spin fluctuations (see equation (7)). This calibration enables us to quantify the magnitude of magnetization noise and compare it to the saturation magnetization $M_S$ of the $Sm_{0.7}Er_{0.3}FeO_3$ sample. We assume the proportionality of the Faraday rotation angle $\alpha$ with magnetization along the laser propagation direction $M_z$, predicted for magnetized solids [26], and find the equation:

$$\alpha(t) = \underbrace{\mu_0 \cdot V(\lambda) \cdot d}_{\sqrt{C_M}} \cdot M_z(t)$$

(33)

Here, the Verdet constant $V(\lambda)$ at wavelength $\lambda$, the sample thickness $d$ and the vacuum permeability $\mu_0$ define the proportionality factor $C_M$. Typically, literature represents the magneto-optic constant as $V' = V(\lambda)\mu_0 M_S$ (in deg/cm), which assumes saturated magnetization and does not allow direct conclusions about instantaneous magnetization in A/m.

To determine $C_M$ and $V(\lambda)$, we require both the oop magnetization component $M_z$ along the laser propagation axis $z$ and the corresponding polarization rotation $\alpha$. We measure $M_z$ with superconducting quantum interference device (SQUID) magnetometry and compare it to Faraday rotation data obtained at similar temperatures and external fields. Measurements span a range of temperatures through the spin reorientation transition [27] (SRT) in orthoferrite $Sm_{0.7}Er_{0.3}FeO_3$. Here, the spin lattice continuously rotates by 90° with temperature. The SRT initiates at $T_L$ (where $\boldsymbol{M}||a$) and completes at $T_U$ (where $\boldsymbol{M}||c$) along the material's crystalline axes $a$ and $c$ (FIG 6a).

SQUID magnetometry measurements were performed on a cylindrical bulk sample of $Sm_{0.7}Er_{0.3}FeO_3$ (radius $r = 3.1$ mm, thickness $d = 1.7$ mm) grown by the floating-zone technique. Details of the measurements are provided in Appendix G, and data is plotted in FIG 6b. For the optical measurements, we use a 10 μm thick single-crystal platelet from the same batch as the bulk sample. The laser is aligned along the $c$-axis, with a $\pm 62$ mT magnetic field applied in the same direction. The FR data is shifted by +7 K to correct for laser-induced heating [17] and symmetrization removes the linear background attributed to sample birefringence (see Appendix G). The resulting curves are presented in FIG 6b.

Above 313 K, both FR and SQUID magnetometry show a similar trend: the FR angle and measured oop magnetization continuously increase with temperature, saturating at $T_U \approx 334$ K. This is consistent with spin reorientation from the $a$-axis to the $c$-axis (FIG 6a). We denote the saturation magnetization as $M_S := M(T_U) \approx 8.4\ \frac{\text{kA}}{\text{m}}$. Below 313 K, the FR angle trends toward zero as expected from the SRT, though SQUID data shows finite magnetization. This is likely due to sample misalignment. Furthermore, it needs to be noted that SQUID measures the spatially averaged magnetization of the whole sample, while FR probes the local environment around the focus spot. This may also lead to discrepancies between the two methods.

We determine $C_M$ by computing the ratio between FR and SQUID data for all temperatures above 313 K. Following equation (33), the resulting ratios are averaged and squared to obtain:

*Contact author: marvin.weiss@uni-konstanz.de
†Contact author: takayuki.kurihara@issp.u-tokyo.ac.jp

$$C_M = (9.9 \pm 3.2)\frac{(\mu\text{rad})^2}{(\text{A/m})^2} \qquad (34)$$

Using $C_M$, the Verdet constant at 771 nm is:

$$V(771\text{ nm}) = \frac{\sqrt{C_M}}{\mu_0 d} = (2.5 \pm 0.8)\cdot 10^5 \frac{\text{rad}}{\text{T}\cdot\text{m}} \qquad (35)$$

With $M_S \approx 8.4\,\frac{\text{kA}}{\text{m}}$, the normalized Verdet constant $V'$ in $\text{Sm}_{0.7}\text{Er}_{0.3}\text{FeO}_3$ is approximately 1510 deg/cm, comparable to values for $\text{SmFeO}_3$ ($V' \approx 800$ deg/cm) and $\text{ErFeO}_3$ ($V' \approx 2300$ deg/cm) in Ref [28].

### D. Quantitative magnetization noise

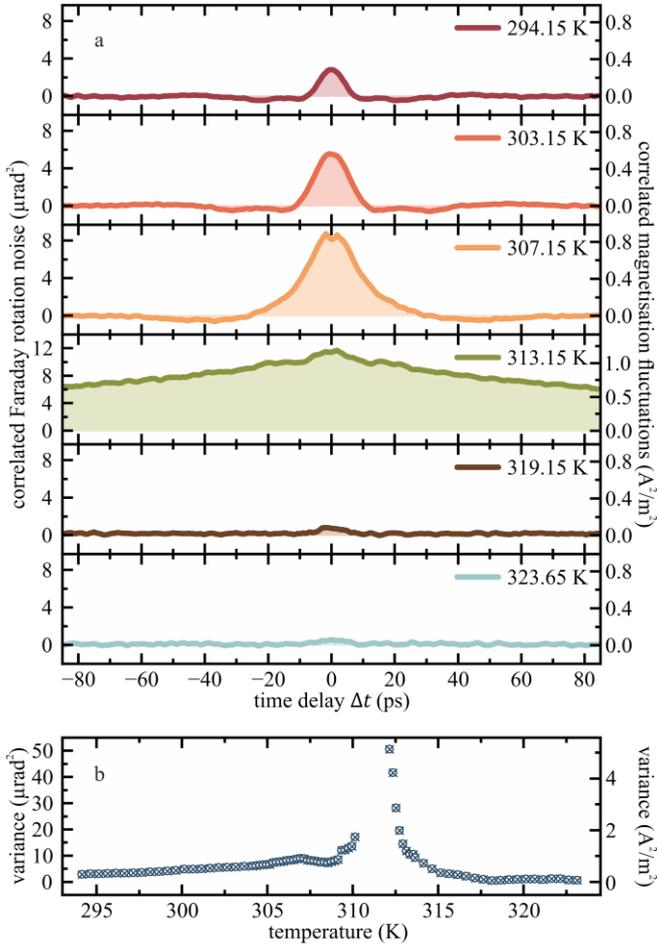

FIG 7: Ultrafast spin noise near the spin reorientation in $\text{Sm}_{0.7}\text{Er}_{0.3}\text{FeO}_3$ calibrated in absolute units of $(\mu\text{rad})^2$ and $\frac{\text{A}^2}{\text{m}^2}$. The data presented here is taken from Ref. [17] and calibrated using the protocols provided in the main text. a Correlated noise dynamics as a function of the probe-probe time delay $\Delta t$ for multiple temperatures near the spin reorientation in $\text{Sm}_{0.7}\text{Er}_{0.3}\text{FeO}_3$. b Time-zero amplitude (variance) as a function of temperature.

Finally, we apply our calibration to the experimental data presented in our previous work (Ref. [17]). Here, correlated spin fluctuations in $\text{Sm}_{0.7}\text{Er}_{0.3}\text{FeO}_3$ were investigated near the SRT. The calibrated correlation waveforms, depicted in FIG 7a, were measured using a probing power of 1.13 mW. These measurements span from -85 ps to +85 ps in 0.6 ps increments and represent the average of 5 - 30 repetitions each. The variance, defined as the correlation amplitude at $\Delta t = 0$, and the temporal dynamics of the noise correlation exhibit significant temperature dependence around the SRT. FIG 7b illustrates the variance as a function of temperature. As the temperature approaches 307 K, a steady increase in variance is observed. This increase is linked to the anisotropy softening around the SRT which results in an enhanced magnetic susceptibility. Beyond this temperature, a sharp increase in variance around 312 K is found, resulting from ultrafast spontaneous spin switching [17]. This phenomenon manifests as an additional exponentially decaying feature, which is superimposed on the damped oscillation resulting from the correlated fluctuations of the quasiferromagnetic magnon mode (green curve, FIG 7a). For a detailed discussion of the temporal dynamics, refer to Ref. [17]. At higher temperatures, the noise amplitude continuously decreases, nearly vanishing around 320 K. This behavior is consistent with the reduction of the oop fluctuation component due to the equilibrium rotation of the spin system within the SRT (see FIG 6a).

*Contact author: marvin.weiss@uni-konstanz.de

†Contact author: takayuki.kurihara@issp.u-tokyo.ac.jp

At the lowest investigated temperature of 294.15 K, we find the magnitude of the correlated magnon noise to reach approximately $0.3\,\frac{\text{A}^2}{\text{m}^2}$. Here, contributions of the picosecond random telegraph switching are assumed to be negligible. From approximately 310 K, picosecond switching contributes significantly to the correlated spin noise, reaching variances as large as $5.1\,\frac{\text{A}^2}{\text{m}^2}$. These values show that magnitude of oop magnon fluctuations and picosecond switching constitute less than 0.01 % and 0.03 % of the saturation magnetization amplitude $M_S$, respectively. Here, we referenced the standard deviation (square-root of variance) of the magnetization noise to $M_S$. It is important to note that the spin noise represents spatially averaged magnetization fluctuations within the probing volume. This fact leads to significant variance differences when altering the laser focusing, as reported in Ref [17]. At more microscopic spatial scales, the amplitude of local spin fluctuations is expected to exceed the calibrated values presented above.

## IV. CONCLUSION

In summary, we present three calibration methods to quantitatively link a femtosecond noise correlation signal to correlated polarization noise in femtosecond probe pulse trains. All methods yield consistent results, thereby validating our calibration approaches. By combining these methods, we determine a combined calibration factor $C_{\alpha,\text{AVE}}(P_{0,\text{ref}}) = 4.6 \pm 0.8\,\frac{\mu\text{V}}{\mu\text{rad}^2}$ for our current setup. Additionally, we estimate the sensitivity of the setup as $S_{\text{rad}} = 0.52\,\frac{\mu\text{rad}^2}{\text{mW}\sqrt{n}}$ per measurement repetition $n$, facilitating tailored measurements for specific samples. Furthermore, by comparing Faraday rotation experiments with SQUID magnetometry, we calibrate the magnetization fluctuations in absolute units of $\frac{\text{A}^2}{\text{m}^2}$. Applying these calibration techniques to FemNoC data captured in $\text{Sm}_{0.7}\text{Er}_{0.3}\text{FeO}_3$ within the SRT regime, we find that the magnon fluctuations are on the order of 0.01 % and ultrafast spontaneous switching reaches 0.03 % of the saturation magnetization. This quantitative evaluation of the spin noise magnitude in an antiferromagnet provides important information for the development of future spintronic devices.

## ACKNOWLEDGEMENTS

This research was supported by the Japan Society for the Promotion of Science (JSPS) KAKENHI (Grant Nos. JP21K14550, JP24H00317, JP24H02232, and JP23K17748), and the Deutsche Forschungsgemeinschaft (DFG, German Research Foundation) - Project No. 425217212-SFB 1432.

## COMPETING INTERESTS

The authors declare no competing interests.

## DATA AVAILABILITY

The datasets generated in this study, along with the corresponding raw data and analysis files are publicly available (Ref. [29]).

## APPENDIX A: Shot noise in a balanced photodetection system:

We investigate the fluctuations in the optical power difference $\Delta P = P_s - P_p$ between s- and p-polarized components, as measured in a balanced photodetector. The variance of $\Delta P$ is given by:

$$\langle (\Delta P)^2 \rangle = \langle (P_s - P_p)^2 \rangle = \langle P_s^2 + P_p^2 - 2P_s P_p \rangle$$
$$= \langle P_s^2 \rangle + \langle P_p^2 \rangle - 2\langle P_s P_p \rangle$$

(A1)

where $\langle P_{s,p}^2 \rangle$ represents the variance of s- and p-polarization components, and $\langle P_s P_p \rangle$ is their cross-correlation. First, we analyze the individual variances:

$$\langle P_{s,p}^2 \rangle = \langle (\langle P_{s,p} \rangle + \delta P_{s,p})^2 \rangle$$
$$= \langle \langle P_{s,p} \rangle^2 + \delta P_{s,p}^2 + 2\langle P_{s,p} \rangle \delta P_{s,p} \rangle$$
$$= \langle P_{s,p} \rangle^2 + \langle (\delta P)_{s,p}^2 \rangle + 2\langle P_{s,p} \rangle \langle \delta P_{s,p} \rangle$$
$$= \langle P_{s,p} \rangle^2 + \langle (\delta P)_{s,p}^2 \rangle$$

(A2)

Here, $\langle P_{s,p} \rangle$ is their mean power, and $\delta P_{s,p}$ represents their fluctuations, where $\langle \delta P_{s,p} \rangle = 0$. Assuming Poissonian statistics (with $\langle n^2 \rangle = \langle n \rangle$ (Ref. [30])), and using the definition for the mean optical power $\langle P \rangle = \frac{h\nu}{\tau}$ averaged in the time window $\tau$, the fluctuations in power are:

$$\langle (\delta P)_{s,p}^2 \rangle := \langle (\delta P)_{s,p}^2 \rangle_{SN} = 2h\nu \langle P_{s,p} \rangle \Delta f$$

(A3)

where $h$ is Planck's constant, $\nu$ is the photon frequency and $\Delta f = \frac{1}{\tau}$ is the noise bandwidth. The cross-correlation term $\langle P_s P_p \rangle$ simplifies as:

$$\langle P_s P_p \rangle = \langle (\langle P_s \rangle + \delta P_s) \cdot (\langle P_p \rangle + \delta P_p) \rangle$$
$$= \langle \langle P_s \rangle \langle P_p \rangle + \langle P_s \rangle \delta P_p + \delta P_s \langle P_p \rangle + \delta P_s \delta P_p \rangle$$
$$= \langle \langle P_s \rangle \langle P_p \rangle \rangle + \langle P_s \rangle \langle \delta P_p \rangle + \langle \delta P_s \rangle \langle P_p \rangle + \langle \delta P_s \delta P_p \rangle$$

(A4)

Again, we use that the mean of the fluctuation components is zero $\langle \delta P_{s,p} \rangle = 0$. Furthermore, we assume the fluctuation components in s- and p-polarization components to be uncorrelated ($\langle \delta P_s \delta P_p \rangle = 0$), which simplifies equation (A4) to:

$$\langle P_s P_p \rangle = \langle \langle P_s \rangle \langle P_p \rangle \rangle = \langle P_s \rangle \langle P_p \rangle$$

(A5)

By combining equations (A2), (A3) and (A5) into (A1), we obtain:

$$\langle (\Delta P)^2 \rangle = (\langle P_s \rangle^2 + \langle (\delta P)_s^2 \rangle) + (\langle P_p \rangle^2 + \langle (\delta P)_p^2 \rangle)$$
$$- 2\langle P_s \rangle \langle P_p \rangle$$
$$= \langle P_s \rangle^2 + \langle P_p \rangle^2 + 2h\nu \Delta f (\langle P_s \rangle + \langle P_p \rangle)$$
$$- 2\langle P_s \rangle \langle P_p \rangle$$

(A6)

Using Malus' law [20], $\Delta P = P_s - P_p = [P_0 \sin^2(\alpha)] - [P_0 \cos^2(\alpha)] = P_0(1 - 2\cos^2(\alpha))$, where $\alpha$ is the input beam's polarization angle, we simplify the terms of equation (A6):

$$\langle P_s \rangle^2 + \langle P_p \rangle^2 - 2\langle P_s \rangle \langle P_p \rangle$$
$$= \langle P_0 \sin^2(\alpha) \rangle^2 + \langle P_0 \cos^2(\alpha) \rangle^2$$
$$- 2\langle P_0 \sin^2(\alpha) \rangle \langle P_0 \cos^2(\alpha) \rangle$$
$$= P_0^2 [\sin^4(\alpha) + \cos^4(\alpha) - 2\sin^2(\alpha)\cos^2(\alpha)]$$
$$= P_0^2 \cos^2(2\alpha) = \langle \Delta P \rangle^2$$

(A7)

Inserting (A7) into (A6) then yields:

*Contact author: marvin.weiss@uni-konstanz.de
†Contact author: takayuki.kurihara@issp.u-tokyo.ac.jp

$$\langle(\Delta P)^2\rangle = \langle\Delta P\rangle^2 + \underbrace{2h\nu P_0 \Delta f}_{\langle\delta\Delta P^2\rangle}$$

(A8)

This shows that the shot noise fluctuations $\langle(\delta\Delta P)^2\rangle = \langle(\delta P_s - \delta P_p)^2\rangle$ in the power difference are equivalent to the shot noise level of a single beam with optical power $P_0$:

$$\langle(\Delta P)^2\rangle_{SN} = 2h\nu P_0 \Delta f = \langle(\delta P_s - \delta P_p)^2\rangle$$

(A9)

## APPENDIX B: Variance of pulse-to-pulse fluctuation:

A key aspect of femtosecond noise correlation spectroscopy is the extraction of pulse-to-pulse polarization fluctuations, $\Delta\delta\alpha$, using subharmonic lock-in detection. This method not only removes stationary background signals [18] but also doubles the noise variance compared to $\delta\alpha$. Below is a mathematical proof for this variance doubling:

Consider two independent random variables $X$ and $X'$, independently drawn from the same probability distribution $p(X)$ with expectation value $\mu[X]$ and variance $V(X) = \mu(X^2) - \mu(X)^2$. We calculate the variance of their difference $X - X'$:

$$\begin{aligned}V(X - X') &= \mu[(X-X')^2] - \mu[X-X']^2\\ &= \mu[X^2 + X'^2 - 2XX']\\ &\quad - (\mu[X] - \mu[X'])^2\\ &= \mu[X^2] + \mu[X'^2] - 2\mu[XX']\\ &\quad - (\mu[X]^2 + \mu[X']^2 - 2\mu[X]\mu[X'])\\ &= \underbrace{\mu[X^2] - \mu[X]^2}_{V(X)} + \underbrace{\mu[X'^2] - \mu[X']^2}_{V(X')}\\ &\quad - 2\underbrace{(\mu[XX'] - \mu[X]\mu[X'])}_{\text{cov}(X,X')}\end{aligned}$$

(A10)

Here, $\text{COV}(X, X')$ is the covariance between $X$ and $X'$. Since $X$ and $X'$ are independent, their covariance is zero, and because $X$ and $X'$ and follow identical distributions, $V(X) = V(X')$. Thus, we have:

$$V(X - X') = 2V(X)$$

(A11)

This shows that the variance of the difference between two uncorrelated random variables is twice the variance of each individual variable. This applies to femtosecond noise correlation spectroscopy, where subharmonic lock-in detection doubles the noise variance and must be accounted for in calibration. Note: This is valid only when polarization noise in consecutive pulses is uncorrelated. In our case, the 50 ns interpulse distance from our 40 MHz laser system is much larger than the sub-nanosecond noise correlation time reported in Ref. [17].

*Contact author: marvin.weiss@uni-konstanz.de

†Contact author: takayuki.kurihara@issp.u-tokyo.ac.jp

## APPENDIX C: Post processing of the correlation data:

The data presented in FIG 3, FIG 5 and FIG 11 undergo several post-processing steps, described below. The data in FIG 7 are reproduced from Ref. [17]. Corresponding data processing procedures can be found in that reference.

### 1. Processing of data in FIG 3

The raw correlation data (solid lines in FIG 8) for the $\langle(\Delta\delta V_1 \pm \Delta\delta V_2)^2\rangle$ curves in FIG 3 exhibit a strong linear background drift and a sharp peak at $\Delta t = 0$. The drift likely results from slight misalignment of the optical delay line, while the peak arises from spectral leakage – i.e., a small fraction of one probe beam entering the other detection path due to imperfect separation by the dichroic mirror. This leads to pulse interference within the same detection branch.

Each raw dataset is averaged over 10 measurement repetitions, with time delays ranging from $-80$ ps to $+80$ ps in 600 fs steps.

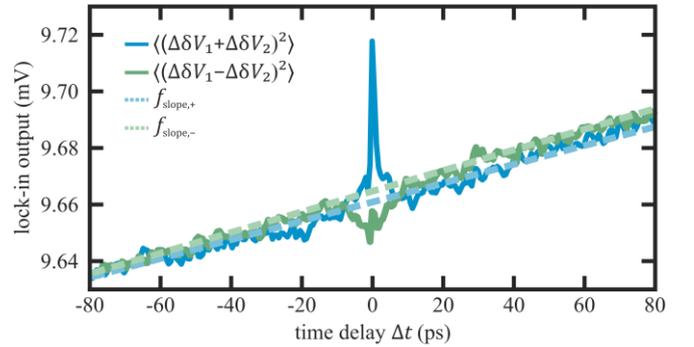

FIG 8: Slope correction of the raw correlation data according to the procedure described in the text. The solid lines represent the raw data $\langle(\Delta\delta V_1 \pm \Delta\delta V_2)^2\rangle$ and the dashed lines their corresponding slope functions $f_{\text{slope},\pm}$.

To correct for the linear drift and obtain slope-corrected data $\langle(\Delta\delta V_1' \pm \Delta\delta V_2')^2\rangle_{\text{sc}}$, we apply the following transformation:

$$\begin{aligned}\langle(\Delta\delta V_1 \pm \Delta\delta V_2)^2\rangle_{\text{sc}} &= \langle(\Delta\delta V_1 \pm \Delta\delta V_2)^2\rangle - f_{\text{slope}}(\Delta t)\\ &\quad + f_{\text{slope}}(\Delta t = 0)\end{aligned}$$

(A12)

Here, $f_{\text{slope}}(\Delta t)$ is defined as the average of the slope functions computed for both the sum and difference signals:

$$f_{\text{slope}}(\Delta t) = \frac{1}{2}\sum_\pm (\Delta t - \Delta t_{i=1})S_\pm + \langle(\Delta\delta V_1 \pm \Delta\delta V_2)^2\rangle_{i\in[1,5]}$$

(A13)

The slope $S_\pm$ is calculated by dividing the difference between the average of the last and first 5 data points by the total time window:

$$S_\pm = \frac{\langle(\Delta\delta V_1 \pm \Delta\delta V_2)^2\rangle_{i\in[N-5,N]} - \langle(\Delta\delta V_1 \pm \Delta\delta V_2)^2\rangle_{i\in[1,5]}}{|\Delta t_N - \Delta t_1|}$$

(A14)

where, $i \in \mathbb{Z}\{0\}$ indexes the data points and $N$ is the total number of points. The slope functions $f_{\text{slope},\pm}(\Delta t)$ are shown as dashed lines in FIG 8, and the corrected data appear as solid lines in FIG 3.

After slope correction, we remove the interference artifact at $\Delta t = 0$ by omitting a 1.2 ps centered at this point. Finally, a third-order Savitzky-Golay filter is applied to smooth the data, resulting in the dotted curves shown in FIG 3.

2. Processing of data in FIG 5 and FIG 11

The raw correlated noise waveforms shown in FIG 5 and FIG 11 are first slope-corrected using equation (A12), then offset-subtracted to remove the linear offset arising in the lock-in detector. The slope-corrected waveforms are fitted with the function:

$$f(\Delta t) = c + A \cdot \text{sech}\left(\frac{\Delta t}{\tau}\right) \cos(2\pi \Delta t f)$$

(A15)

where $\tau$ is the damping timescale, $f$ is the oscillation frequency, and $A$ and $c$ are the amplitude and offset, respectively. The final curves shown in FIG 5 and FIG 11 are obtained by subtracting the fitted offset $c$. No data points were omitted, and no Savitzky-Golay filtering was applied in this case.

## APPENDIX D: Calibration via acousto-optic modulator:

In this calibration method, we introduce a controllable subharmonic (20 MHz) amplitude modulation with optical power $P_{20}$ on the probe beams (total power $P_0$) using an acousto-optic modulator (AOM). Since this modulation affects both the s- and p-polarized components equally, the subharmonic power difference $\Delta P_{20}$ depends on the mean polarization angles $\langle \alpha_{1,2} \rangle$ (see equation ( 21 )).

Directly determining $P_{20}$ and $\Delta P_{20}$ is challenging, so instead, we calculate the ratio of the balanced photodetector (BPD) output voltage variances $\frac{\langle V_{\text{BPD}}^2(f_{\text{rep}})\rangle}{\langle V_{\text{BPD}}^2\left(\frac{f_{\text{rep}}}{2}\right)\rangle} = \frac{P_0^2}{P_{20}^2}$ (see equation ( 26 )) at both the laser's fundamental frequency $f_{\text{rep}}$ and its subharmonic $\frac{f_{\text{rep}}}{2}$, using the spectrum analyzer function of our lock-in amplifier. FIG 4b shows an example power spectrum recorded at power $P_0 = 1.09$ mW and polarization angles $\langle \alpha_1 \rangle = 205°$ and $\langle \alpha_2 \rangle = 195°$ for detection channels 1 and 2. The 40 MHz and 20 MHz peak amplitudes are extracted from the spectra recorded for different angles $\langle \alpha \rangle$, with the results shown in FIG 9a.

Note that we plot the absolute angles of the half-wave plates (HWP), which may have an offset from the polarization angle. The 40 MHz and 20 MHz components show minima at different polarization angles, which we attribute to the BPD's spectral response. To enhance statistics, we interpolate the data in FIG 9a using the fitting function:

*Contact author: marvin.weiss@uni-konstanz.de
†Contact author: takayuki.kurihara@issp.u-tokyo.ac.jp

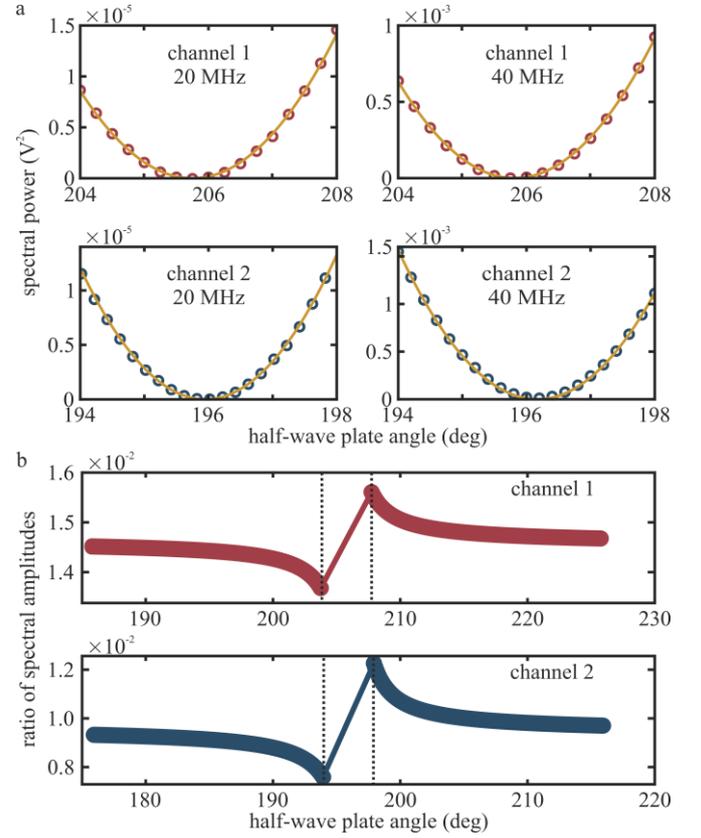

FIG 9: Estimation of spectral amplitude ratio of 20 MHz and 40 MHz components. a. Spectral power of 20 MHz (left) and 40 MHz peaks (right) as a function of the absolute half-wave plate angles of detection channels 1 (red circles) and 2 (blue circles). A square fit (yellow) is used to interpolate the measured values. b. Ratio between spectral amplitudes of the 20 MHz and 40 MHz components for detection channels 1 (red circles) and 2 (blue circles) calculated from the fits in a. Due to an angle mismatch of the minimal spectral amplitudes of 20 MHz and 40 MHz components, the ratio is not constant as a function of the half-wave plate angle. To avoid artifacts arising from this mismatch, we neglect the data points between the dashed lines in the following discussion.

$$\langle (\Delta V_{20,40})^2 \rangle (\langle \alpha \rangle)$$
$$= \Bigg( c \cdot \left(1 - 2\cos^2\left(2\pi \cdot \frac{\langle \alpha \rangle - \langle \alpha_0 \rangle}{360°} - \frac{\pi}{4}\right)\right) \Bigg)^2$$
$$\approx 4c^2 \left(\frac{2\pi(\langle \alpha \rangle - \langle \alpha_0 \rangle)}{360°} - \frac{\pi}{4}\right)^2$$

(A16)

Here, $\langle \alpha_0 \rangle$ is the offset between half-wave plate (HWP) angle and the 45° polarization reference. Note that the small angle approximation used in equation ( 21 ) is justified in this case, because $\langle \alpha \rangle$ is varied by less than 5°.

From the fits in FIG 9a, we infer the ratio of spectral amplitudes as a function of the HWP angles (FIG 9b). While the ratio $\frac{\langle V_{\text{BPD}}^2(f_{\text{rep}})\rangle}{\langle V_{\text{BPD}}^2\left(\frac{f_{\text{rep}}}{2}\right)\rangle}$ is expected to remain constant, a slight angle dependency is observed due to minima occurring at different angles. To address this, we discard ratio data within

2° of the subharmonic minimum (dashed lines, FIG 9b) and average the remaining angles, yielding the mean ratio of $\frac{\langle V_{\text{BPD}}^2(f_{\text{rep}})\rangle}{\langle V_{\text{BPD}}^2(\frac{f_{\text{rep}}}{2})\rangle} = 0.012 \pm 0.004$.

To extract the proportionality factor $D$ for calibration, we examine the lock-in response to the mean polarization angles $\langle \alpha_{1,2} \rangle$ of the 20 MHz amplitude-modulated beam. We simultaneously vary the HWP angles of detection channels 1 and 2 relative to the fitted minima in FIG 9a from -0.25° to +0.29° in 0.02° steps and measure the lock-in correlation output. The data, shown in FIG 10, follows the quadratic behavior with a quadratic coefficient $D = 0.102 \pm 0.003 \frac{\text{V}}{(\mu\text{rad})^2}$ obtained from the polynomial fit. A small linear background persists, likely due to imperfections in the detection branches, such as finite extinction ratios of the HWPs or nonlinearities in the photo detectors.

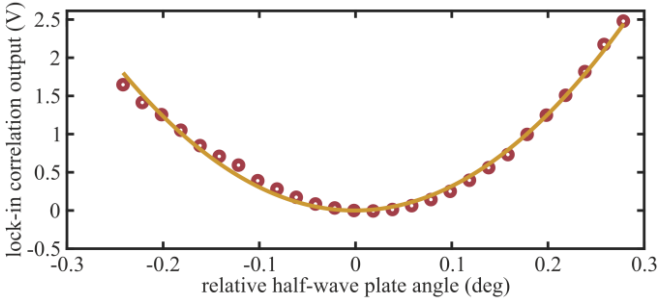

FIG 10: Estimation of the proportionality factor $D$ relating the lock-in correlation output to the mean polarization angles $\langle \alpha_{1,2} \rangle$ of the 20 MHz amplitude modulated probe beams. A polynomial fit (yellow) with fitting function $f(x) = cx + Dx^2$ yields $c = 0.072 \pm 0.102 \frac{\text{V}}{\mu\text{rad}}$ and $D = 0.102 \pm 0.003 \frac{\text{V}}{(\mu\text{rad})^2}$.

## APPENDIX E: Detailed quantification of measurement values and uncertainty evaluation:

In this section, we summarize all experimental parameters, their uncertainties (denoted as $u(A)$, where $A$ is the experimental parameter), and their sources (whether estimated, calculated, or fitted). Unless stated otherwise, uncertainties are calculated using Gaussian error propagation [31]:

$$u(f(x_i)) = \sqrt{\sum_i \left(\frac{\partial f}{\partial x_i} \cdot u(x_i)\right)^2}$$

(A17)

Here, $u(f(x_i))$ is the combined uncertainty of the function $f(x_i)$, which depends on various parameters $x_i$, each with its own uncertainties $u(x_i)$.

The values and their uncertainties used in the individual calibration schemes (transfer function method - section IIIa1, optical shot noise comparison - section IIIa2, and acousto-optic modulator - section IIIa2) are summarized in Table 1, Table 2, and Table 3, respectively.

*Contact author: marvin.weiss@uni-konstanz.de
†Contact author: takayuki.kurihara@issp.u-tokyo.ac.jp

Following these calibration schemes, we determine the individual calibration factors $C_{\alpha,\text{TF}}$, $C_{\alpha,\text{SN}}$ and $C_{\alpha,\text{AOM}}$, along with their uncertainties $u(C_{\alpha,\text{TF}})$, $u(C_{\alpha,\text{SN}})$ and $u(C_{\alpha,\text{AOM}})$. From these values, a combined calibration factor $C_{\alpha,\text{AVE}}$ is calculated with corresponding uncertainty $u(C_{\alpha,\text{AVE}})$. This is done using the inverse-variance weighted mean method [23], which assigns greater weight to values with smaller uncertainties, effectively accounting for the experimental errors of each calibration method:

$$C_{\alpha,\text{AVE}} = \frac{\sum_{i \in \{\text{TF,SN,AOM}\}} \frac{C_i}{u(C_i)^2}}{\sum_{i \in \{\text{TF,SN,AOM}\}} \frac{1}{u(C_i)^2}}$$

$$= 4.60 \frac{\mu\text{V}}{(\mu\text{rad})^2} \quad \text{and} \quad u(C_{\text{AVE}})$$

$$= \sqrt{\frac{1}{\sum_{i \in \{\text{TF,SN,AOM}\}} \frac{1}{u(C_i)^2}}} = 0.77 \frac{\mu\text{V}}{(\mu\text{rad})^2}$$

(A18)

At last, the setup is calibrated in terms of magnetization in the units of $\frac{\text{A}^2}{\text{m}^2}$. The experimental values used in this calibration scheme, as well as their uncertainties are summarized in Table 4.

## APPENDIX F: Sensitivity of the FemNoC system:

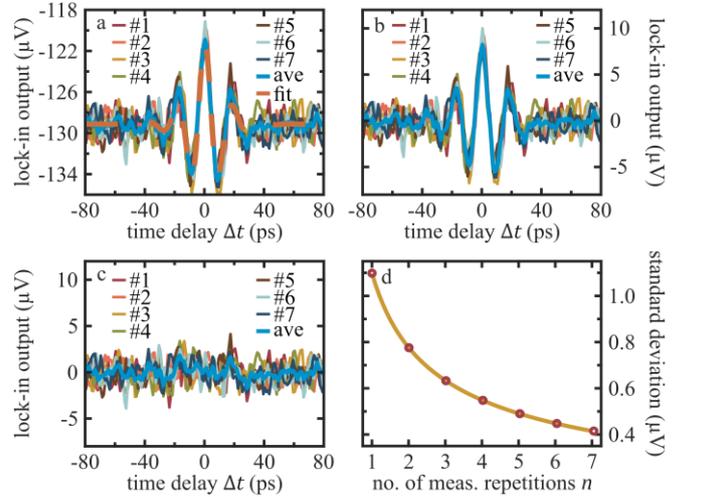

FIG 11: a. Slope corrected data. b. Slope corrected and background subtracted data. c. Background noise extracted via subtraction of fit from waveforms in a. d. Average standard deviation of determined background noise as function of waveform averages (red circles) and inverse square root fit (yellow line).

We estimate the Faraday noise sensitivity of the FemNoC system as a function of probing power $P_0$ and the number of measurement repetitions $n$. This is based on the standard deviation ($STD$) of the background noise in the measured correlation noise time traces. Seven waveforms of correlated magnetization noise were recorded for $Sm_{0.7}Er_{0.3}FeO_3$ 297 K under $63 \pm 5$ mT out-of-plane magnetic field, using probe power $P_{0,\text{ref}} = 1.09$ mW, with data points from $\Delta t = -80$ ps to $\Delta t = +80$ ps in 1 ps steps. Each data point represents an average of approximately $10^7$ correlation samples. FIG 11a shows the slope-corrected waveforms, their average, and a fit using function (A15).

The offset-subtracted time traces are shown in FIG 11b, and the remaining background noise (obtained by subtraction of the fit from slope-corrected waveforms) is plotted in FIG 11c. The $STD$ of the first and last 20 ps of the time traces was calculated to avoid artifacts stemming from deviations of fit and measured data. We repeated this calculation for all possible combinations of the seven recorded waveforms, averaging them $n$ at a time, and plotted the results in FIG 11d. As expected, the data follows an inverse square-root trend with averaging.

To generalize the results for different optical powers, we use the scaling $STD \propto P_0$ (Ref. [18]) and $STD \propto \frac{1}{\sqrt{n}}$ (FIG 11d), giving the generalized $STD$:

$$STD(P_0, n) = STD(P_{0,\text{ref}}, n_{\text{ref}}) \cdot \frac{P_0}{P_{0,\text{ref}}} \cdot \sqrt{\frac{n_{\text{ref}}}{n}}$$

(A19)

The sensitivity of our setup $S(P_0, n)$ is defined as twice the standard deviation of the background noise: $S(P_0, n) \stackrel{\text{def}}{=} 2 \cdot STD(P_0, n)$. To convert this to Faraday rotation noise sensitivity $S_{\text{rad}}(P_0, n)$ units of $(\mu\text{rad})^2$, we divide by the average calibration factor $C_{\text{AVE}}(P_0)$ (equation (29)):

$$S_{\text{rad}}(P_0, n) = \frac{S(P_0, n)}{C_{\text{AVE}}(P_0)} = \frac{2 STD(P_{0,\text{ref}}, n_{\text{ref}})}{C_{\text{AVE}}(P_{0,\text{ref}})} \cdot \frac{P_{0,\text{ref}}}{P_0} \cdot \sqrt{\frac{n_{\text{ref}}}{n}}$$
$$\stackrel{\text{def}}{=} \frac{S_{\text{rad,ref}}(P_{0,\text{ref}}, n_{\text{ref}})}{P_0 \sqrt{n}}$$

(A20)

Using $P_0 = P_{0,\text{ref}} = 1.09$ mW, $n = n_{\text{ref}} = 1$, and $STD(P_{0,\text{ref}}, 1) = 1.1$ μV (FIG 11d), we calculated the sensitivity normalized to 1 mW of probe power:

$$S_{\text{rad}}(P_0, n) = \frac{S_{\text{rad}}(1.09 \text{ mW}, 1)}{P_0 \sqrt{n}} = 0.52 \frac{\mu\text{rad}^2}{\text{mW}\sqrt{n}}$$

(A21)

This provides the sensitivity per measurement repetition with the current setup, though it can be generalized for any number of correlation samples per time-delay data point.

**APPENDIX G: Post-processing of Faraday rotation data:**

The Faraday rotation (FR) data in Figure 7 undergoes several post-processing steps. Raw data was recorded under external out-of-plane fields of $-63$ mT and $+61$ mT, similar to the fields used in SQUID magnetometry. Furthermore, we measure FR curves at small external fields of $-5$ mT and $+4$ mT. The average of values recorded below 297 K from the small-field data serves as the zero reference level, with all FR curves shown in FIG 12a.

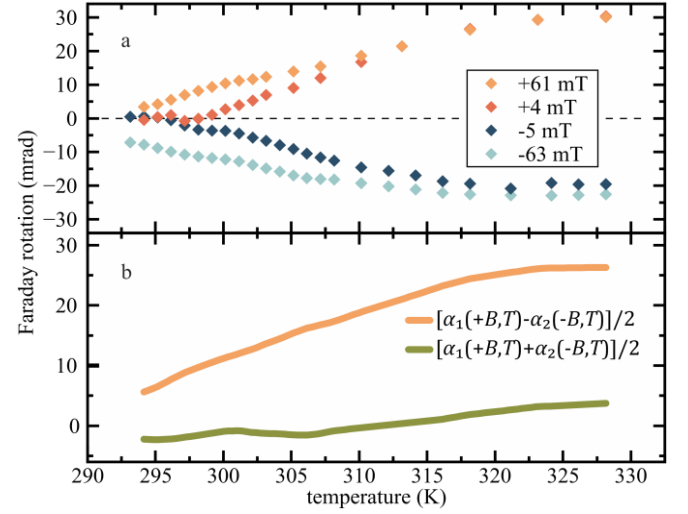

FIG 12: a. Raw Faraday rotation as a function of temperature. All rotation angles are referenced to average values recorded below 297 K from the small-field data ($-5$ mT, $+4$ mT). b. Difference (yellow) and sum (green) of the $+61$ mT and $-63$ mT FR curves. The sum reveals a linear background with temperature, which we assign to birefringence in the orthoferrite sample [32].

The absolute FR angles differ slightly between positive and negative field pairs of similar magnitude. This effect may be explained by birefringence of the sample, which also changes with temperature [32]. To separate the background, we calculate the sum and difference of the $-63$ mT and $+61$ mT curves. The results are depicted in FIG 12b. The sum (green curve) isolates the background, while the difference (yellow curve) provides the background-free magneto-optic response.

*Contact author: marvin.weiss@uni-konstanz.de
†Contact author: takayuki.kurihara@issp.u-tokyo.ac.jp

*Contact author: marvin.weiss@uni-konstanz.de

†Contact author: takayuki.kurihara@issp.u-tokyo.ac.jp

Table 1: Experimental values and uncertainties of calibration via transfer function method.

| Name | Symbol | Value | Method of estimation | Uncertainty | Method of error determination |
|---|---|---|---|---|---|
| Optical probe power | $P_0$ | 1.09 mW | Optical power meter | 0.05 mW | Estimated |
| Transfer function 1 | $C_1$ | 2.2 mW | Calculated: $C_1 = 2P_0$ | 0.1 mW | Combined error |
| Responsivity | $R(\lambda = 771 \text{ nm})$ | $0.53 \frac{A}{W}$ | Specification sheet of Thorlabs BPD 450A-AC (Ref. [19]) | $0.08 \frac{A}{W}$ | Estimated |
| Amplitude frequency response @ $G = 10^5 \frac{V}{A}$; $f = 20 \text{ MHz}$ | $s(G, f)$ | $-11.37 \text{ dB} = 0.27 \frac{A}{W}$ | Specification sheet of Thorlabs BPD 450A-AC (Ref. [19]) | $1.5 \text{ dB} = 0.02 \frac{A}{W}$ | Estimated |
| Transimpedance gain | $G$ | $10^5 \frac{V}{A}$ | Specification sheet of Thorlabs BPD 450A-AC (Ref. [19]) | $10^3 \frac{V}{A}$ | Estimated |
| Transfer function 2 | $C_2$ | $1.4 \cdot 10^4 \frac{V}{W}$ | Calculated: $C_2 = R(\lambda) \cdot s(G, f) \cdot G$ | $0.3 \cdot 10^4 \frac{V}{W}$ | Combined error |
| Transfer function 3 | $C_3$ | $4.95 \cdot 10^3 \frac{1}{V}$ | Fit | $0.05 \cdot 10^3 \frac{1}{V}$ | 95% error bounds of fit |
| Calibration factor – transfer function | $C_{\alpha,\text{TF}}$ | $4.8 \frac{\mu V}{(\mu rad)^2}$ | Calculated: $C_{\alpha,\text{TF}} = C_1^2 C_2^2 C_3$ | $1.6 \frac{\mu V}{(\mu rad)^2}$ | Combined error |

*Contact author: marvin.weiss@uni-konstanz.de
†Contact author: takayuki.kurihara@issp.u-tokyo.ac.jp

Table 2: Experimental values and uncertainties of calibration via optical shot noise.

| Name | Symbol | Value | Method of estimation | Uncertainty | Method of error determination |
|---|---|---|---|---|---|
| Optical probe power | $P_0$ | 1.09 mW | Optical power meter | 0.05 mW | Estimated |
| Centre wavelength | $\lambda$ | 771 nm | Optical spectrum analyzer, mean of probe pulses | 5 nm | Estimated |
| Photon energy | $h\nu$ | $2.58 \cdot 10^{-19}$ J | Calculated: $h\nu = \frac{hc}{\lambda}$ | $0.02 \cdot 10^{-19}$ J | Combined error |
| Effective bandwidth of el. bandpass filter | $\Delta f$ | 4 MHz | Specification sheet of Mini-Circuits BBP-21.4+ (Ref. [33]) | 1 MHz | Estimated |
| Total noise level | $\langle(\Delta\delta V)^2\rangle_{tot}$ | $9.7 \cdot 10^{-3}$ V | Calculated from slope-corrected curves (see FIG 3b and FIG 8) | $0.1 \cdot 10^{-3}$ V | Estimated |
| Dark noise level | $\langle(\Delta\delta V)^2\rangle_{dark}$ | $5.1 \cdot 10^{-3}$ V |  | $0.1 \cdot 10^{-3}$ V | Estimated |
| FR noise correlation amplitude | $\langle\Delta\delta V_1 \cdot \Delta\delta V_2(\Delta t = 0)\rangle_{FR}$ | $1.10 \cdot 10^{-5}$ V | $\text{Mean}(\text{Max}(\langle(\Delta\delta V_1 \pm \Delta\delta V_2)^2\rangle_{smooth} - \langle\Delta\delta V^2\rangle_{tot}))$ | $0.06 \cdot 10^{-5}$ V | Estimated |
| FR noise of detection channels 1 and 2 | $\langle(\Delta\delta V_{1,2})^2\rangle_{FR}$ | $2.19 \cdot 10^{-5}$ V | $\langle\Delta\delta V^2\rangle_{FR} = 2\langle\Delta\delta V_1 \cdot \Delta\delta V_2(\Delta t = 0)\rangle_{FR}$ | $0.11 \cdot 10^{-5}$ V | Combined error |
| Shot noise level | $\langle(\Delta\delta V)^2\rangle_{SN}$ | $4.57 \cdot 10^{-3}$ V | $\langle\Delta\delta V^2\rangle_{SN} = \langle\Delta\delta V^2\rangle_{tot} - \langle\Delta\delta V^2\rangle_{dark} - \langle\Delta\delta V_{1,2}^2\rangle_{FR}$ | $0.14 \cdot 10^{-3}$ V | Combined error |
| Calibration factor – shot noise | $C_{\alpha,SN}$ | $4.8 \, \frac{\mu V}{(\mu rad)^2}$ | $C_{\alpha,SN} = \frac{\langle\Delta\delta V^2\rangle_{SN}}{h\nu \Delta f} \cdot \frac{P_0}{rad^2}$ | $1.2 \, \frac{\mu V}{(\mu rad)^2}$ | Combined error |

*Contact author: marvin.weiss@uni-konstanz.de
†Contact author: takayuki.kurihara@issp.u-tokyo.ac.jp

Table 3: Experimental values and uncertainties of calibration via acousto-optic modulator.

| Name | Symbol | Value | Method of estimation | Uncertainty | Method of error determination |
|---|---|---|---|---|---|
| Optical probe power | $P_0$ | 1.09 mW | Optical power meter | 0.05 mW | Estimated |
| Mean spectral amplitude ratio of 40 MHz and 20 MHz components | $\dfrac{\langle V_{\text{BPD}}^2(f_{\text{rep}}) \rangle}{\langle V_{\text{BPD}}^2\left(\frac{f_{\text{rep}}}{2}\right) \rangle}$ | 0.012 | Mean of all calculated ratios in FIG 9b | 0.004 | Standard deviation of all calculated ratios |
| Proportionality factor | $D$ | $0.102 \dfrac{\text{V}}{(\mu\text{rad})^2}$ | Polynomial fit (see FIG 10) | $0.003 \dfrac{\text{V}}{(\mu\text{rad})^2}$ | 95% error bounds of fit |
| Calibration factor – AOM | $C_{\alpha,\text{AOM}}$ | $4.2 \dfrac{\mu\text{V}}{(\mu\text{rad})^2}$ | $C_{\alpha,\text{AOM}} = \dfrac{D}{2} \dfrac{\langle V_{\text{BPD}}^2(f_{\text{rep}}) \rangle}{\langle V_{\text{BPD}}^2\left(\frac{f_{\text{rep}}}{2}\right) \rangle}$ | $1.3 \dfrac{\mu\text{V}}{(\mu\text{rad})^2}$ | Combined error |

Table 4: Experimental values and uncertainties of magnetization calibration.

| Name | Symbol | Value | Method of estimation | Uncertainty | Method of error determination |
|---|---|---|---|---|---|
| FR sample thickness | $d$ | 10 μm | Estimated | 1 μm | Estimated |
| Mean FR/SQUID ratio | $\left\langle \dfrac{\alpha(T)}{M_c(T)} \right\rangle_T$ | $3.2 \dfrac{\mu\text{rad}}{(\text{A/m})}$ | See Appendix G | $1.0 \dfrac{\mu\text{rad}}{(\text{A/m})}$ | Estimated |
| Calibration factor – Magnetization | $C_M$ | $9.9 \dfrac{(\mu\text{rad})^2}{(\text{A/m})^2}$ | $C_M = \left\langle \dfrac{\alpha(T)}{M_c(T)} \right\rangle_T^2$ | $3.2 \dfrac{(\mu\text{rad})^2}{(\text{A/m})^2}$ | Combined Error |
| Verdet constant | $V(771 \text{ nm})$ | $2.5 \cdot 10^5 \dfrac{\text{rad}}{\text{T} \cdot \text{m}}$ | $V(771 \text{ nm}) = \dfrac{\sqrt{C_M}}{\mu_0 d}$ | $0.8 \cdot 10^5 \dfrac{\text{rad}}{\text{T} \cdot \text{m}}$ | Combined Error |

*Contact author: marvin.weiss@uni-konstanz.de
†Contact author: takayuki.kurihara@issp.u-tokyo.ac.jp